\documentclass[letterpaper,titlepage,11pt]{article}
\usepackage{hyperref}
\usepackage{amssymb,amsmath,amsfonts}
\usepackage{epsfig}
\def\clock{{\count0=\time
           \divide\count0 60
           \ifnum\count0<10 0\fi\the\count0
           \multiply\count0 -60 \advance\count0 \time
           :\ifnum\count0<10 0\fi \the\count0
         }}
\newcommand{\timestamp}{{\small\vbox{\hbox{\tt\jobname.tex}
\hbox{\the\day/\the\month/\the\year, \clock}}}}

\newcommand{\CF}{\mathcal{F}}

\newcommand{\CN}{\mathcal{N}}
\newcommand{\CO}{\mathcal{O}}

\newcommand{\R}{\mathbb{R}}

\newcommand{\spa}{\ , \ \ }

\newcommand{\ads}{\mbox{AdS}}

\numberwithin{equation}{section}

\begin{document}

\begin{titlepage}

\rightline{\vbox{\small\hbox{\tt NORDITA-2012-5} }}
 \vskip 1.8 cm

\centerline{\Huge \bf Thermal string probes in AdS}
\vskip 0.3cm
\centerline{\Huge \bf and finite temperature Wilson loops}
\vskip 1.5cm

\centerline{\large {\bf Gianluca Grignani$\,^{1}$},  {\bf Troels Harmark$\,^{2}$},}
\vskip 0.2cm \centerline{\large  {\bf Andrea Marini$\,^{1}$} ,  {\bf Niels A. Obers$\,^{3}$} and
{\bf Marta Orselli$\,^{3}$} }

\vskip 1.0cm

\begin{center}
\sl $^1$ Dipartimento di Fisica, Universit\`a di Perugia,\\
I.N.F.N. Sezione di Perugia,\\
Via Pascoli, I-06123 Perugia, Italy
\vskip 0.4cm
\sl $^2$ NORDITA\\
Roslagstullsbacken 23,
SE-106 91 Stockholm,
Sweden \vskip 0.4cm
\sl $^3$ The Niels Bohr Institute, Copenhagen University  \\
\sl  Blegdamsvej 17, DK-2100 Copenhagen \O , Denmark
\end{center}
\vskip 0.6cm

\centerline{\small\tt grignani@pg.infn.it, harmark@nordita.org, }
\centerline{\small\tt andrea.marini@fisica.unipg.it, obers@nbi.dk, orselli@nbi.dk}

\vskip 1.3cm \centerline{\bf Abstract} \vskip 0.2cm \noindent
We apply a new description of thermal fundamental string probes to the
study of finite temperature Wilson loops in the context of the AdS/CFT
correspondence. Previously this problem has been considered using
extremal probes even though the background is at finite temperature. 
The new description of fundamental string probes is based on the blackfold approach.
As a result of our analysis we find a new term in the potential
between static quarks in the symmetric representation which for sufficiently small temperatures is the leading correction to the Coulomb force potential. We also find an order $1/N$ correction  to the onset of the Debye screening of the quarks.
These effects arise from including the thermal excitations of the string probe and demanding the probe to be in thermodynamic equilibrium with the Anti-de Sitter black hole background.

\end{titlepage}


\section{Introduction}

Wilson loops are an important class of non-local observables in the study of strongly coupled gauge theories. 
Their expectation values provide non-trivial information on confinement, quark screening and phase transitions and they can in particular be used to compute the potential for a quark-antiquark pair. 

This paper is about the holographic description of Wilson loops for finite temperature gauge theory in the planar limit. In the AdS/CFT correspondence a Wilson loop is dual to a fundamental string (F-string) probe with its world-sheet extending into the bulk of Anti-de Sitter space (AdS) and ending at the location of the loop on the boundary of AdS
\cite{Rey:1998ik}. At zero temperature, the classical Nambu-Goto (NG) action provides an effective description of the F-string probe.%
\footnote{This has led to an impressive weak-strong coupling interpolation of circular Wilson loops 
\cite{Erickson:2000af}.}
However, for finite temperature gauge theory the F-string is probing a finite temperature space-time such as an AdS black hole. 
In this case, the classical NG action does not take into account the thermal excitations of the string that arise since one can locally regard a static string probe as being in a heat bath. 
For an AdS black hole this can make the classical NG action an increasingly inaccurate description as one approaches the event horizon since by the Tolman law the local temperature for a static string probe is redshifted towards infinity. Thus, one can possibly miss not just quantitative corrections but also qualitative effects if one uses the classical NG action.

One way to find an accurate description of an F-string probe in a finite temperature background would be to quantize the NG action and in this way include the thermalization of the string degrees of freedom. Indeed, in 
\cite{deBoer:2008gu} the leading quadratic correction was taken into account for a string probing an AdS black hole. However, to find the equivalent of the NG action for finite temperature string probes one should find an effective action that includes all orders in the temperature \cite{Brandhuber:1998bs} (see also the review \cite{Sonnenschein:1999if}). 
In this work we shall describe and apply a new approach to find the effective action for an F-string probe in a finite temperature background. In this approach we use the supergravity (SUGRA) solution for a non-extremal F-string to describe a finite temperature Wilson loop by employing the blackfold methods developed in  
\cite{Emparan:2009cs,Emparan:2009at}%
\footnote{Ref. \cite{Emparan:2007wm} discusses the first application to neutral black rings in asymptotically flat space 
and large classes of new neutral black objects in asymptotically flat and $(A)dS$ backgrounds were
found in  \cite{Emparan:2009vd} and \cite{Caldarelli:2008pz} respectively.  For reviews see \cite{Emparan:2009zz}.  See also the recent paper \cite{Camps:2012hw} for a more general derivation of the blackfold effective
theory.}
and generalized to supergravities including those relevant to string theory in \cite{Emparan:2011hg,Caldarelli:2010xz,Grignani:2010xm}. 
Using this approach one finds an effective description for an F-string probe at finite temperature in which the thermal excitations of the string are fully taken into account. Since the SUGRA F-string requires having many coincident F-strings the finite-temperature F-string probe corresponds to a Wilson loop in the $k$-symmetric product of the fundamental representation where $k$ is the number of F-strings. While the classical NG action is a valid description of a string probe at zero temperature and for $k=1$, $g_s \ll 1$ the new approach is valid for both finite and zero temperature and in the regime $1 \ll k \ll N$, $g_s^2 k \gg 1$. The new approach to string probes parallels the case of finite-temperature D-brane probes and
its application to generalize the BIon solution to finite temperature \cite{Grignani:2010xm,Grignani:2011mr}.

To illustrate the consequences of the new holographic description of finite temperature Wilson loops we consider the specific case of rectangular Wilson loops in finite-temperature $\CN=4$ super Yang-Mills theory with gauge group $SU(N)$ in the large $N$ limit. The expectation value of this Wilson loop gives the potential for a $Q$-$\bar{Q}$ pair in the gauge theory, where $Q$ corresponds
to the symmetric representation of $k$ quarks. This is furthermore related to the tension of $k$-strings.\footnote{The tension of $k$-strings  has been studied in QCD both analytically as well as on the lattice (see e.g. 
\cite{Shifman:2005eb}).}
On the holographically dual side, the gauge theory in question is described as a black hole in $\ads_5$.
This is in the Poincar\'e patch corresponding to the fact that the superconformal gauge theory is deconfined.  The $Q$-$\bar{Q}$ pair is dual to $k$ coincident
probe F-strings attached to the locations of the two particles at the boundary of $\ads_5$. 

One important general motivation of studying quark-gluon plasma physics using the AdS/CFT correspondence is that $\CN=4$ SYM theory at finite temperature shares many similarities with QCD, with the advantage that the former is
much easier to study. Consequently, the study of Wilson loops at finite temperature and their application to
the physics of quark-antiquark pairs, has provided an impetus to study dynamical issues, such
as the energy loss of a quark moving through a strongly coupled plasma (see e.g.  \cite{Herzog:2006gh}).  
In many instances such studies provide a very good approximation to the physics of QCD including
Debye screening in strongly coupled non-abelian plasma.  For example, motivated by the observation that at weak coupling $\CN=4$ SYM theory approximates very well the physics of QCD if one compares the two theories for coinciding values of the Debye mass~\cite{Huot:2006ys}, in \cite{Bak:2007fk} it was examined to which extent the similarity between the two theories holds at strong coupling.%
\footnote{For the numerous interesting applications, recent developments and other works,
we refer the reader to the review \cite{CasalderreySolana:2011us} which  explains and emphasizes the important role of using AdS/CFT techniques to interpret the physics of QCD and the phenomenology of heavy ion collisions.}

Our analysis reveals various new results and we highlight the main points here. Denoting the distance between the
$Q$-$\bar{Q}$ pair by $L$, we find in the small $LT$ expansion of the free energy for the rectangular
Wilson loop a {\it new correction term} to the Coulomb potential,  as compared to the first correction 
observed using the extremal F-string probe. In particular, for sufficiently small temperatures the former is the leading 
correction to the Coulomb force potential (see Eq.~\eqref{FloopLT}). We also study the rectangular Wilson loop for finite values of $LT$. As expected, we find that there is a phase transition to a phase with two Polyakov loops, one for each charge, corresponding to the Debye screening of the charges. However, we find an order $1/N$ correction to the onset of the Debye screening relative to what an extremal F-string probe gives. 
In addition to these results we perform a careful analysis of the validity of the probe approximation and show that it 
breaks down close to the event horizon.

\section{Thermal F-string probe}
\label{sec:probe}

The blackfold approach 
\cite{Emparan:2009cs,Emparan:2009at,Emparan:2011hg,Caldarelli:2010xz,Grignani:2010xm} 
 is a general framework that enables one to describe black branes in the probe approximation. The probe approximation means that the thickness of the brane is much smaller than any of the length scales in the background or in the embedding of the brane (apart from the length scale corresponding to the local temperature). In the case of a thermal F-string probe we employ the SUGRA solution of $k$ coincident black F-strings in type IIB SUGRA in 10D Minkowski space. From this we can compute the energy-momentum tensor $T_{00} = A r_0^6 ( 7 + 6 \sinh^2\alpha )$, $T_{11} = - A r_0^6 (1+6\sinh^2\alpha)$ with $A= \Omega_7 / (16\pi G)$. The temperature is $3 / ( 2\pi r_0 \cosh \alpha)$ and the charge is $k T_{\rm F1} = 6 A r_0^6 \cosh \alpha \sinh \alpha $ with $T_{\rm F1} = 1/ (2\pi l_s^2)$. As found in \cite{Emparan:2009at,Emparan:2011hg}  the action principle for a stationary blackfold uses the free energy as the action since the first law of thermodynamics is equivalent to the EOMs. For a SUGRA F-string probe in a general background with redshift factor $R_0$ one finds the free energy 
\begin{equation}
\CF = A \int dV_{(1)} R_0 r_0^6 ( 1 + 6 \sinh^2 \alpha) 
\end{equation}
Note that for this free energy the action principle requires holding the charge $k$ fixed during a variation.

The background we want to probe is the AdS black hole in the Poincar\' e patch times a five-sphere. This is because we want to study the rectangular Wilson loop in $\CN=4$ SYM on $S^1 \times \R^3$, the $S^1$ meaning that it is at finite temperature. The metric of this background is
\begin{equation}
\label{adsbhmet}
ds^2 = \frac{R^2}{z^2} ( - f dt^2 + dx^2 + dy_1^2 + dy_2^2 + f^{-1} dz^2 ) + R^2 d\Omega_5^2 \spa f(z)=1-\frac{z^4}{z_0^4}
\end{equation}
where $R$ is the AdS radius, the boundary of AdS is at $z=0$ and the event horizon is at $z=z_0$. The temperature of the black hole as measured by an asymptotic observer is $T = 1 / (\pi z_0)$.
Note that the five-form Ramond-Ramond flux will not play any role in the following. We consider now the following ansatz for the embedding of the F-string probe
\begin{equation}
\label{ansatz}
t = \tau \equiv \sigma^0 \spa z = \sigma \equiv \sigma^1 \spa x=x(\sigma)
\end{equation}
We take this ansatz since we want to study a static string probe that extends between the point charge $Q$ at the location $(x,y_1,y_2)=(0,0,0)$ and the point charge $\bar{Q}$ at the location $(x,y_1,y_2)=(L,0,0)$ on the boundary of AdS at $z=0$. Furthermore, the string is located at a point on the five-sphere which on the gauge theory side means that we are breaking the R-symmetry. See also the discussion in Section \ref{sec:concl} for further comments on this.
The induced metric and redshift factor for this embedding are
\begin{equation}
\gamma_{ab} d\sigma^a d\sigma^b = \frac{R^2}{\sigma^2} \left[ - f 
d\tau^2 + \Big( f^{-1}+ x'{}^2 \Big)  d\sigma^2  \right] \spa R_0( \sigma) = \frac{R}{\sigma} \sqrt{f(\sigma)}
\end{equation}
It is important to note that the redshift factor $R_0$ induces a local temperature $T_{\rm local} = T /R_0$ which is the temperature that the static string probe locally is subject to, $T$ being the global temperature of the background space-time as 
measured by an asymptotic observer. Thus imposing $T /R_0 = 3 / ( 2\pi r_0 \cosh \alpha)$ ensures that the probe is in thermal equilibrium with the background. Taking everything into account the free energy for the thermal F-string probe in the background \eqref{adsbhmet} with the ansatz for the embedding \eqref{ansatz} is
\begin{equation}
\label{free_energy}
\CF = A  \left( \frac{3}{2\pi T} \right)^6 \int d\sigma \sqrt{1 + f(\sigma) x'(\sigma)^2} G(\sigma)  \spa G(\sigma) \equiv \frac{R^8}{\sigma^8} f(\sigma)^3  \frac{1+6\sinh^2 \alpha(\sigma)}{\cosh^6 \alpha (\sigma)}
\end{equation}
Varying with respect to $x(\sigma)$ this gives the EOM
\begin{equation}
\label{eomF1}
\left( \frac{ f(\sigma)x'(\sigma) }{\sqrt{1+ f(\sigma) x'(\sigma){}^2}} G(\sigma) \right)' = 0
\end{equation}
For the most part we will be considering solutions with $x' > 0$ and the boundary conditions $x(0)=0$ and $x' \rightarrow \infty$ for $\sigma \rightarrow \sigma_0$. In this case one can write the general solution as
\begin{equation}
\label{sol1}
x' (\sigma) = \left( \frac{f(\sigma)^2 G(\sigma)^2}{f(\sigma_0) G(\sigma_0)^2} - f(\sigma) \right)^{-\frac{1}{2}} 
\end{equation}
The full solution which goes from $(x,z)=(0,0)$ to $(x,z) = (L/2,\sigma_0)$ and back to $(x,z)=(L,0)$ is found by mirroring the solution \eqref{sol1} around $x=L/2$.
The distance $L$ between the charge $Q$ and the charge $\bar{Q}$ is then
\begin{equation}
\label{theL}
L = 2 \int_0^{\sigma_0} d\sigma \left( \frac{f(\sigma)^2 G(\sigma)^2}{f(\sigma_0) G(\sigma_0)^2} - f(\sigma) \right)^{-\frac{1}{2}} 
\end{equation}
For clarity we introduce the dimensionless parameter $\hat{\sigma} = \pi T \sigma$. In terms of this we find
\begin{equation}
\label{LTeq}
L T = \frac{2}{\pi} \int_0^{\hat{\sigma}_0} d\hat{\sigma} \left( \frac{f(\hat{\sigma})^2 H(\hat{\sigma})^2}{f(\hat{\sigma}_0) H(\hat{\sigma}_0)^2} - f(\hat{\sigma}) \right)^{-\frac{1}{2}} 
\end{equation}
with $f(\hat{\sigma}) = 1 - \hat{\sigma}^4$, $\hat{\sigma}_0 = \pi T \sigma_0$, 
\begin{equation}
\label{kappa}
 H(\hat{\sigma}) = \frac{f(\hat{\sigma})^3}{\hat{\sigma}^8}   \frac{1+6\sinh^2 \alpha(\hat{\sigma})}{\cosh^6 \alpha (\hat{\sigma})}
\spa
\kappa  \equiv  \frac{2^5 k T_{\rm F1}}{3^7 A R^6} = \frac{f(\hat{\sigma})^3}{\hat{\sigma}^6} \frac{ \sinh \alpha(\hat{\sigma})}{\cosh^5 \alpha (\hat{\sigma})}
\end{equation}
Thus, we see that $LT$ depends only on the dimensionless quantities $\kappa$ and $\hat{\sigma}_0$. Note that the equation for $\kappa$ enforces the charge conservation and can be used to find $\alpha(\hat{\sigma})$. In terms of the gauge theory variables $k$, $\lambda$ and $N$, $\kappa$ can be written
\begin{equation}
\label{gaugekappa}
\kappa = \frac{2^7}{3^6} \frac{k \sqrt{\lambda}}{N^2}
\end{equation}
using the AdS/CFT dictionary for the AdS radius $R^4 = \lambda l_s^4$ and the string coupling $4\pi g_s = \lambda / N$ with $\lambda$ being the 't Hooft coupling of $\CN=4$ SYM theory.

In addition to the above solution of the EOM \eqref{eomF1} we consider in the following also another type of solution,
namely that of an F-string extending from a point at the boundary straight down towards the horizon. This
solution trivially solves \eqref{eomF1} with $x' (\sigma)= 0$. In detail we consider the two solutions 
extending from the boundary at either $(z,x)=(0,0)$ or $(z,x)=(0,L)$, $i.e.$ at the position of the $Q$ or the $\bar{Q}$ point charge.

\subsubsection*{Critical distance}

From the equation for $\kappa$ in \eqref{kappa} and the fact that $\sinh \alpha / \cosh ^5 \alpha$ is bounded from above with maximal value $2^4 / 5^{5/2}$ we see that for a given $\kappa$ the equation can only be satisfied provided $\hat{\sigma} \leq \hat{\sigma}_c$ with $\hat{\sigma}_c$ given by
\begin{equation}
\label{hatsigmac}
\hat{\sigma}_c^2 = \sqrt{1 + \frac{5^{5/3}}{2^{14/3}} \kappa^{\frac{2}{3}} } - \frac{5^{5/6}}{2^{7/3}}  \kappa^{\frac{1}{3}}
\end{equation}
We see from this that $\hat{\sigma}_c < 1$ which means that one reaches the critical distance $\hat{\sigma}_c$ before reaching the horizon. The physical origin for this critical distance is that we require the F-string probe to be in thermal equilibrium with the background \eqref{adsbhmet}. The critical distance then arises because the SUGRA F-string has a maximal temperature for a given $k$ and since the Tolman law means that the local temperature goes to infinity as one approaches the black hole. This is a qualitatively new effect which means that the probe description breaks down beyond the critical distance $\hat{\sigma}_c$. We analyze in detail the validity of the probe approximation below.

\subsubsection*{Probe approximation}

It is important to consider the validity of the probe approximation for the SUGRA F-string. In the context of the blackfold approach, the probe approximation essentially means that we should be able to piece the string probe together out of small pieces of SUGRA F-strings in hot flat ten-dimensional space-time. For this to work, the local length scale of the string probe, $i.e.$ the ``thickness" of the string, should be much smaller than the length scale of each of the pieces of F-string in hot flat space. Since we want to consider the branch of the SUGRA F-string connected to the extremal F-string the thickness of the F-string is the charge radius $r_c = r_0 ( \cosh \alpha \sinh \alpha )^{1/6}$, as one can see following similar considerations in 
\cite{Grignani:2010xm,Grignani:2011mr}. Using the formulas above, one finds easily that $r_c \propto \kappa^{1/6} R$. Looking at the AdS black hole background \eqref{adsbhmet} we find that $R$ and $z_0 \propto 1/T$ are the two length scales in the metric. Since we need to require that the sizes of the pieces of F-string should be smaller than the length scales of the metric we need that $r_c \ll R$ and $r_c \ll 1/T$. This gives the conditions
\begin{equation}
\label{kappacond}
\kappa \ll 1 \spa RT \ll \kappa^{- \frac{1}{6}}
\end{equation}
Given that $\kappa \ll 1$, the condition $RT \ll \kappa^{-1/6}$ is easily fulfilled as it just gives a rather weak upper bound on how high the asymptotic temperature $T$ can be. As an immediate consequence of the condition $\kappa \ll 1$ we see that the critical distance $\hat{\sigma}_c$ given in Eq.~\eqref{hatsigmac} - for which the probe reaches the maximal temperature and beyond which the probe description in terms of a SUGRA F-string description breaks down - is very close to the horizon: $1- \hat{\sigma}_c \propto \kappa^{1/3}$ for small $\kappa$. 

As stated earlier, the regime of validity of the SUGRA F-string is $1 \ll k \ll N$ and $\lambda^2 k \gg N^2$. Instead, the probe approximation requires $\kappa \ll 1$. We see that this is consistent with the regime of validity of the SUGRA F-string provided that $\lambda \ll N^2$. This translates to $g_s \ll N$ which is trivially satisfied since we assume weak string coupling $g_s \ll 1$ throughout this paper.

The conditions \eqref{kappacond} are not sufficient to ensure the validity of the probe approximation. Indeed, our analysis of the length scales of the background \eqref{adsbhmet} is only valid away from the near horizon region $z_0 - \sigma \ll z_0$. In the near horizon region $z_0 - \sigma \ll z_0$ further analysis is required. A relevant quantity to consider is the local temperature $T_{\rm local} = T /R_0$. In order for the probe approximation to be valid we need that the local temperature varies over sufficiently large length scales such that we can regard the probe locally as a SUGRA F-string in hot flat space of temperature $T_{\rm local}$. Thus, we need in particular that $r_c T_{\rm local}'(\sigma) /T \ll 1$, $i.e.$ that the variation of the local temperature is small over the length scale of the F-string probe. We compute
\begin{equation}
\label{Tlocalprime}
\frac{r_c T_{\rm local}'}{T} = \frac{r_c}{R \sqrt{f}} + \frac{2 r_c (\pi T \sigma)^4}{R f^{3/2}}
\end{equation}
As a check we can see that for $\sigma \ll z_0$ the condition $r_c T_{\rm local}' /T \ll 1$ reduces to $r_c \ll R$.
Considering instead the near horizon region $z_0 - \sigma \ll z_0$ we find in the $\kappa \rightarrow 0$ limit that $r_c T_{\rm local}' /T \ll 1$ requires $z_0 - \sigma \gg \kappa^{1/9} z_0$. Thus, when we reach $z_0 -\sigma \sim \kappa^{1/9} z_0$, the probe approximation breaks down. In particular we notice that for $\sigma \simeq \sigma_c$ the probe approximation is not valid, indeed $r_c T_{\rm local}' /T \sim \kappa^{-1/3}$.
Thus, the probe approximation in fact breaks down before we reach the critical distance $\sigma=\sigma_c$ where the local temperature reaches the maximal possible temperature of the SUGRA F-string. 

Finally, we should also consider the extrinsic curvature of the solution \eqref{sol1}. Since we already took into account the variation of the background, the easiest way to analyze the extrinsic curvature is to neglect the derivatives of the metric. Doing this we can write the length scale of the extrinsic curvature as
\begin{equation}
L_{\rm ext} (\sigma)= \frac{R (1+f {x'}^2)^{3/2}}{\sigma f|x''|}
\end{equation}
It is not hard to see that the extrinsic curvature is maximal at the turning point $\sigma=\sigma_0$ where we find
\begin{equation}
L_{\rm ext} (\sigma_0) = \frac{2R }{\sigma_0 \sqrt{f(\sigma_0)}} \left|  \frac{f'(\sigma_0)}{f(\sigma_0)} + \frac{2 G'(\sigma_0)}{G(\sigma_0)} \right|^{-1}
\end{equation}
Analyzing this for $\sigma \ll z_0$ we find $L_{\rm ext} (\sigma_0) \sim R$ which means that the probe approximation is valid as long as $r_c \ll R$. Considering instead $\sigma \simeq \sigma_c$ we find $L_{\rm ext} \sim r_c$, thus the probe approximation breaks down at this point, hence we should require $z_0 -\sigma \gg \kappa^{1/3} z_0$. However, this is already guaranteed by the stronger condition $z_0 -\sigma \gg \kappa^{1/9} z_0$ found above.

\subsubsection*{Small $\kappa$ expansion of solution}

Since by \eqref{kappacond} we need to require $\kappa \ll 1$ it is useful to find the solution \eqref{sol1} to the EOM in this limit. We first record that $\alpha(\hat{\sigma})$ from \eqref{kappa} is
\begin{equation}
\label{coshalphaexp}
\cosh^2 \alpha = \frac{(1-\hat{\sigma}^4)^{\frac{3}{2}}}{\hat{\sigma}^3 \sqrt{\kappa}} - \frac{1}{4} - \frac{5}{32} \frac{\hat{\sigma}^3 \sqrt{\kappa}}{(1-\hat{\sigma}^4)^{\frac{3}{2}}} + \CO (\kappa)
\end{equation}
Using this in \eqref{sol1} we find
\begin{eqnarray}
\label{xprimeexp}
x' = \pi T \frac{dx}{d\hat{\sigma}} &=& \frac{\hat{\sigma}^2 \sqrt{1-\hat{\sigma}_0^4}}{\sqrt{(1-\hat{\sigma}^4)(\hat{\sigma}_0^4 - \hat{\sigma}^4 )}} \left( 1 - \frac{\sqrt{\kappa}}{3} \frac{\hat{\sigma}_0^7 \left( \frac{1-\hat{\sigma}^4}{1-\hat{\sigma}_0^4} \right)^{\frac{3}{2}} - \hat{\sigma}^3 \hat{\sigma}_0^4} {\sqrt{1-\hat{\sigma}^4} (\hat{\sigma}_0^4 -\hat{\sigmaÕ}^4 )} + \CO (\kappa) \right)
\end{eqnarray}
We use these formulas below to analyze both the small $\kappa$ limit as well as the small $LT$ limit.

\section{Physics of the rectangular Wilson loop}
\label{sec:results}

\begin{figure}[h!]
\centerline{\includegraphics[scale=0.8]{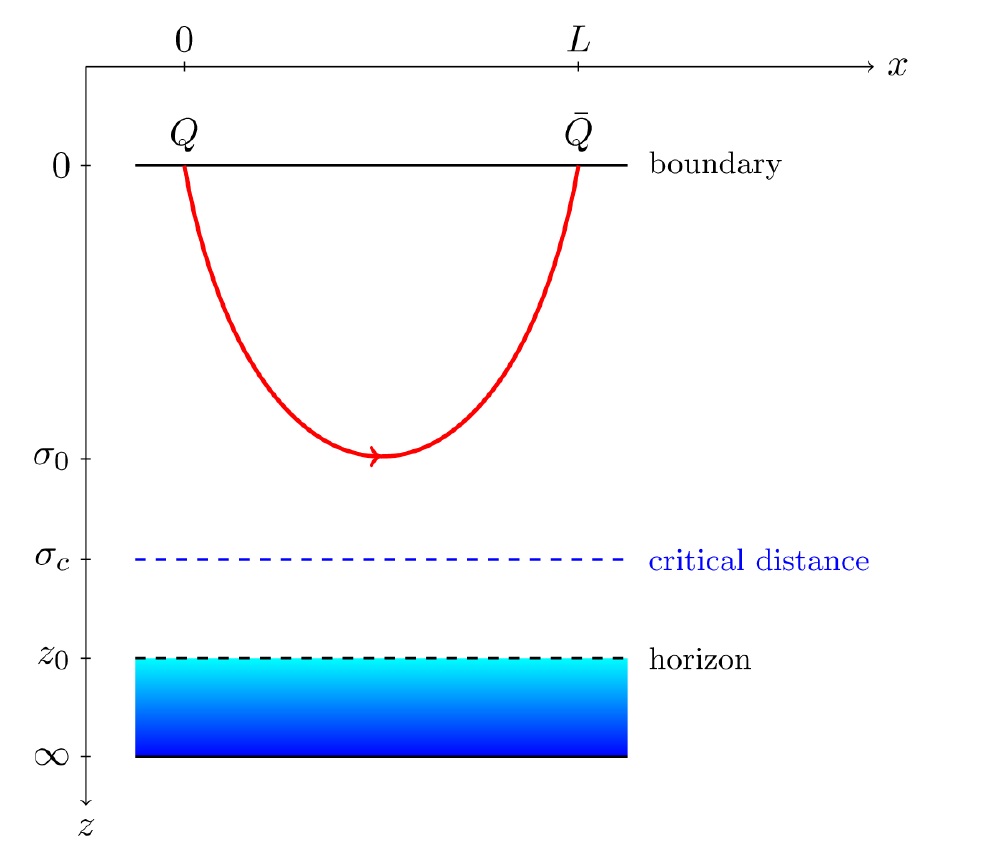}\includegraphics[scale=0.8]{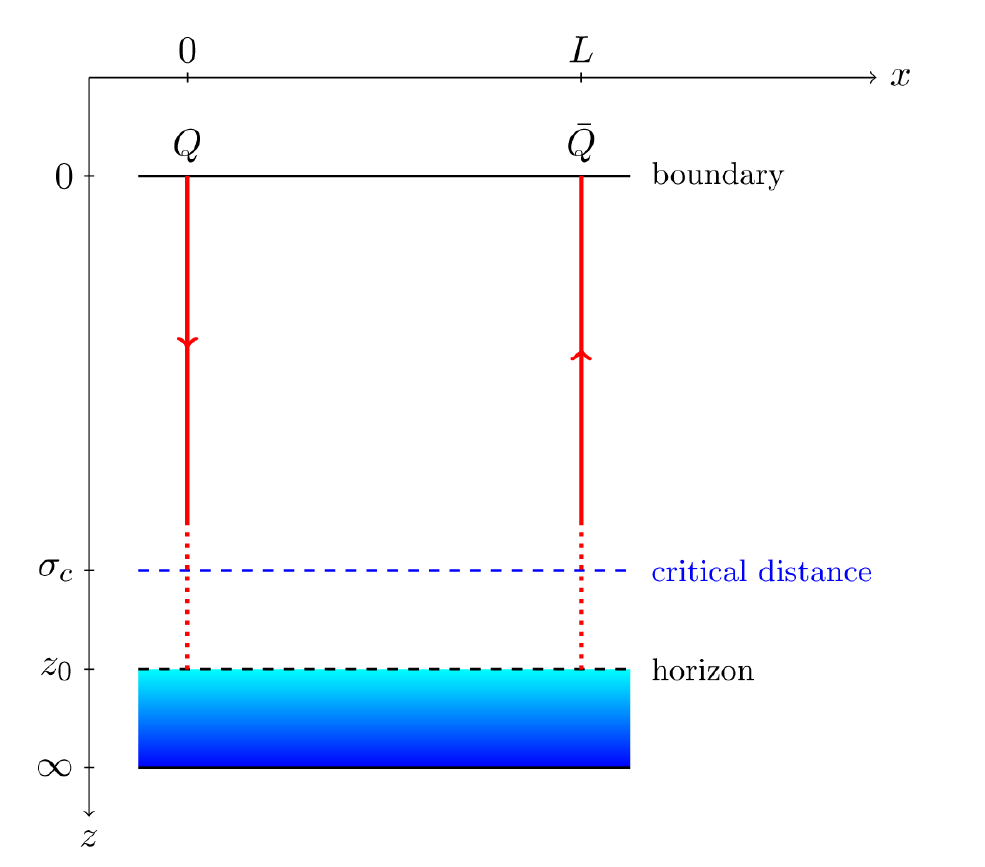}}
	\caption{Left side: F-strings stretched between the $Q$-$\bar{Q}$ pair corresponding to a rectangular Wilson loop. Right side: Two straight strings stretching from the two charges to the event horizon corresponding to two 
	Polyakov loops.}	
		\label{fig:stringprobe}
\end{figure}

We now record our results for a $Q$-$\bar{Q}$ pair in $\CN=4$ SYM at temperature $T$ using the new F-string probe method described above. Our setup is that we have a point charge $Q$ in the $k$-symmetric product of the fundamental representation located at $(z,x)=(0,0)$ and the opposite charge $\bar{Q}$ in the $k$-symmetric product of the anti-fundamental representation located at $(z,x)=(0,L)$. Between the pair of charges we have $k$ coincident 
F-strings probing the AdS black hole background \eqref{adsbhmet}. We have illustrated this in Figure \ref{fig:stringprobe}.

\subsubsection*{Regularized free energy}

From \eqref{free_energy} we can write the free energy of the string extended between the $Q$ and $\bar{Q}$ charges as
\begin{equation}
\label{free_energy2}
\CF = \sqrt{\lambda} k T \int_0^{\hat{\sigma}_0}  \frac{d\hat{\sigma}}{\hat{\sigma}^2}( 1- X )  \sqrt{ 1+ f {x'}^2 } \spa X \equiv  1- \tanh \alpha - \frac{1}{6 \cosh \alpha \sinh \alpha} 
\end{equation}
We see that this is divergent near $\sigma=0$ corresponding to an ultraviolet divergence in the gauge theory. The interpretation of this is that \eqref{free_energy2} is the ``bare" free energy of the rectangular Wilson loop. We thus need a prescription to compute the regularized free energy. The physically most appealing one is to compare the free energy of the rectangular Wilson loop to that of two Polyakov loops, one for each of the point charges $Q$ and $\bar{Q}$. 

In the gauge theory a Polyakov loop is a Wilson line along the thermal circle. For the charge $Q$ ($\bar{Q}$) the Polyakov loop is in the $k$-symmetric product of the fundamental representation (anti-fundamental representation). Seen from the string side, the Polyakov loops corresponds to two strings stretching from the two charges in a direct line towards the event horizon with $x=0$ and $x=L$ 
(as shown in Figure \ref{fig:stringprobe}). This is trivially a solution of the EOM \eqref{eomF1} since $x'=0$. 

The prescription is thus to subtract the ``bare" free energy of the two Polyakov loops since the ultraviolet divergences are the same in this case. To do this in a controlled way, we introduce an infrared cutoff at $z=\sigma_{\rm cut}$ near the event horizon. Note that $\sigma_{\rm cut} \leq \sigma_c$. The ``bare" free energy we should subtract from \eqref{free_energy2} is
\begin{equation}
\label{free_energy3}
\CF_{\rm sub} = \sqrt{\lambda} k T \int_0^{\hat{\sigma}_{\rm cut}}  \frac{d\hat{\sigma}}{\hat{\sigma}^2} ( 1- X )
\end{equation}
The resulting difference $\Delta \CF = \CF - \CF_{\rm sub}$ between the free energies of the rectangular Wilson loop and the two Polyakov loops takes the form
\begin{equation}
\label{DeltaF}
\Delta \CF = \CF_{\rm loop} - 2\CF_{\rm charge}
\end{equation}
with
\begin{equation}
\label{Floop}
\CF_{\rm loop} ( T , L , k , \lambda )= \sqrt{\lambda} k T \left( - \frac{1}{\hat{\sigma}_0} + \int_0^{\hat{\sigma}_0} d\hat{\sigma} \frac{(1-X)\sqrt{1+f {x'}^2} - 1}{\hat{\sigma}^2} \right)
\end{equation}
\begin{equation}
\label{Fcharge}
\CF_{\rm charge} (T , k , \lambda , \sigma_{\rm cut})  = - \frac{1}{2} \sqrt{\lambda} k  T \left( \frac{1}{ \hat{\sigma}_{\rm cut}} + \int_0^{\hat{\sigma}_{\rm cut}} d\hat{\sigma} \frac{X}{\hat{\sigma}^2}
 \right) 
\end{equation}
where $\CF_{\rm loop}$ is interpreted as the regularized free energy of the rectangular Wilson loop and $\CF_{\rm charge}$ the regularized free energy for each of the Polyakov loops, respectively. Notice that we used the dependence of $\sigma_0 (L, T)$ and $\sigma_{\rm cut}$ to separate the contributions for the two types of free energies.
We see that substituting $\sigma_0$ with $\sigma_{\rm cut}$ in \eqref{Floop}, putting $x'=0$ and dividing with two gives \eqref{Fcharge}, which means that the two free energies \eqref{Floop} and \eqref{Fcharge} can be said to be regularized in the same way.

With respect to the Polyakov loop free energy $\CF_{\rm charge}$ we can only use it as the regularized free energy if we can remove the infrared cutoff. This can be done in the $\kappa \rightarrow 0$ limit where $\hat{\sigma}_c \rightarrow 1$. Letting $\sigma_{\rm cut}$ asymptote to $\sigma_c$ we find the free energy
\begin{equation}
\label{Fcharge2}
\CF_{\rm charge} (T , k , \lambda )  = - \frac{1}{2} \sqrt{\lambda} k T 
\end{equation}
Note in particular that by numerical analysis we find that 
\begin{equation}
 \lim_{\kappa \rightarrow 0} \int_0^{\hat{\sigma}_c} d\hat{\sigma} \frac{X (\hat{\sigma} )}{\hat{\sigma}^2}  = 0
 \end{equation}
We notice that the free energy \eqref{Fcharge2} does not depend on how $\sigma_{\rm cut}$ asymptotes to $\sigma_c$. This is encouraging also in view of the fact that the probe approximation is not valid close to $\sigma_c$. 
While \eqref{Fcharge2} is found in the $\kappa \rightarrow 0$ limit we claim that it is also valid for small non-zero $\kappa$. Indeed, from \eqref{Fcharge} we see that the free energy in general should be of the form $\CF_{\rm charge} = - a(\kappa) \sqrt{\lambda} k T /2$ where $a(\kappa)$ is a constant that depends on $\kappa$. Using $d \CF_{\rm charge} = - S_{\rm charge} dT$ we find the entropy $S_{\rm charge} = a(\kappa) \sqrt{\lambda} k/2$. Thus, a dependence on $\kappa$ in $a(\kappa)$ would mean that the entropy as computed from an extremal F-string probe, corresponding to the $\kappa=0$ case, and our thermal F-string probe should be different when applied at zero temperature $T=0$, which obviously does not make sense. Moreover, one would expect an entropy that is extensive in the number of F-strings $k$ at zero temperature, which also would require that $a(\kappa)$ has no dependence on $\kappa$. Hence we conclude that \eqref{Fcharge2} is also valid for $\kappa > 0$. This means that when using $\Delta \CF = \CF_{\rm loop} - 2\CF_{\rm charge}$ below to compare the free energies of the rectangular Wilson loop with that of the two Polyakov loop we use \eqref{Floop} to find the free energy of the rectangular Wilson loop and \eqref{Fcharge2} for the free energy of the Polyakov loop.

\subsubsection*{Free energy of rectangular Wilson loop for small $LT$}

We begin by exploring the length $L$ between the $Q$ and $\bar{Q}$ charges for small $\hat{\sigma}_0 = \pi T \sigma_0$ using the general result \eqref{xprimeexp} for $\kappa \ll 1$. We find
\begin{equation}
\label{LTexp}
LT = \frac{2\sqrt{2\pi}}{\Gamma (\frac{1}{4})^2}  \hat{\sigma}_0 + \left( \frac{\sqrt{2\pi}}{3\Gamma (\frac{1}{4})^2} - \frac{1}{6} \right) \sqrt{\kappa} \hat{\sigma}_0^4 - \frac{2\sqrt{2\pi}}{5\Gamma (\frac{1}{4})^2} \hat{\sigma}_0^5 + \CO( \hat{\sigma}_0^7 )
\end{equation}
We see that small $LT$ is equivalent to small $\hat{\sigma}_0$. The leading term is the same as for an extremal F-string probing the zero temperature $\ads$ background. For the higher order terms we notice here the appearance of the term proportional to $\sqrt{\kappa} \hat{\sigma}_0^4$ in the expansion. In general the expansion of $LT$ has terms of the form $\kappa^{m/2} \hat{\sigma}_0^{3m+4n+1}$. The terms with $\kappa$ marks an important departure from the results obtained using the extremal F-string to probe the AdS black hole background \eqref{adsbhmet}, which here would correspond to setting $\kappa=0$, thus ignoring the thermal excitations of the F-string. 

Considering now the regularized free energy of the rectangular Wilson loop \eqref{Floop} we find using \eqref{coshalphaexp}, \eqref{xprimeexp} and \eqref{LTexp} the expansion
\begin{equation}
\label{FloopLT}
\CF_{\rm loop}= - \frac{\sqrt{\lambda} k}{L} \left(    \frac{4\pi^2}{\Gamma(\frac{1}{4})^4}  + \frac{\Gamma(\frac{1}{4})^4}{96} \sqrt{\kappa} (LT)^3 + \frac{3\Gamma(\frac{1}{4})^4}{160} (LT)^4 + \cdots \right)
\end{equation}
for $LT \ll 1$. Clearly, the leading term is the well-known Coulomb force potential found by probing $\ads_5 \times S^5$ in the Poincar\' e patch with an extremal F-string \cite{Rey:1998ik}.  Considering the higher order terms in the free energy \eqref{FloopLT} we see that for $\kappa=0$ we regain the results found previously in the literature by using an extremal F-string probe in the AdS black hole background \eqref{adsbhmet}. Thus, the higher order term suppressed as $\sqrt{\kappa} (LT)^3$ compared to the leading term in \eqref{FloopLT} is a new term that appears as consequence of including the thermal excitations of the F-string probe, and demanding that the F-string probe is in thermal equilibrium with the background. We can now compare this to the term suppressed as $(LT)^4$ in \eqref{FloopLT}. Since $\kappa \propto k \sqrt{\lambda} /N^2$ we see that the new term is the dominant correction to the Coulomb force potential provided
\begin{equation}
\label{LTcond}
LT \ll \frac{\sqrt{k} \lambda^{1/4}}{N} 
\end{equation}
Thus, we see that for sufficiently small temperatures the leading correction to the Coulomb force potential can only be seen by including the thermal excitations of the F-string probe. This is a very striking consequence of our new thermal F-string probe technique which means that one misses important physical effects by ignoring the thermal excitations of the F-string probe and merely using the extremal F-string as probe of a thermal background. 

\subsubsection*{Finite $LT$ and Debye screening of charges}

\begin{figure}[h!]
\vskip0.2cm
\centerline{\includegraphics[scale=1.2]{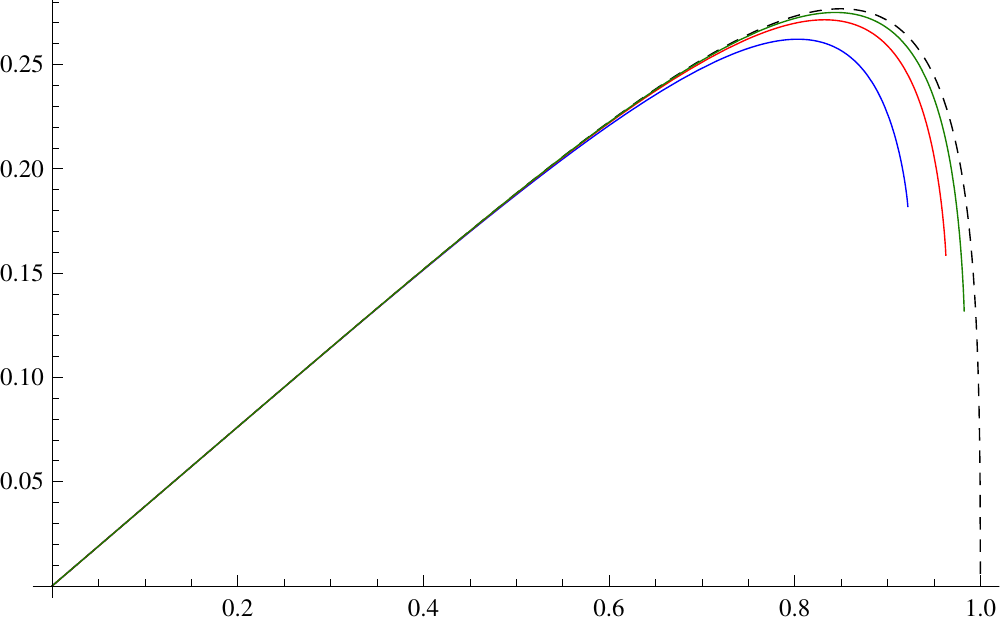}}
\begin{picture}(0,0)(0,0)
\put(28,228){\Large $LT$}
\put(412,27){\Large $\hat{\sigma}_0$}
\end{picture}		
	\caption{
	$LT$ as a function of $\hat{\sigma}_0$ for various values of $\kappa$: 
	$\kappa=0.01$ (blue line), $\kappa=0.001$ (red line), $\kappa=0.0001$ (green line). For comparison we also plot the 
result obtained using the extremal probe (dashed line) \cite{Brandhuber:1998bs} (see Eq.~\eqref{LTsmallkap}). }	
		\label{fig:LTsigmaplots}
\end{figure}
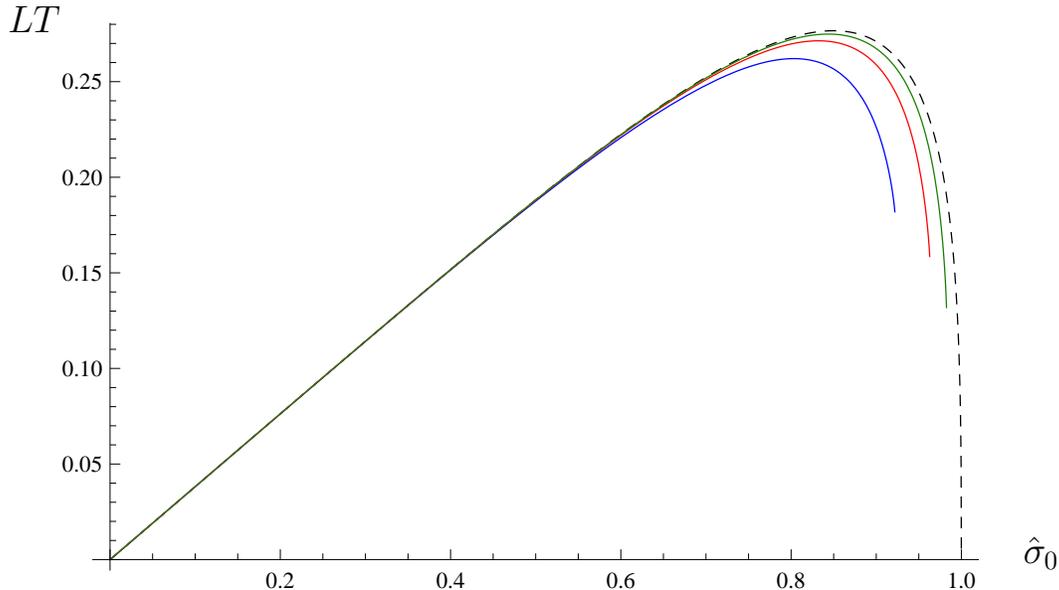

We begin by describing the space of solutions for our static F-string probe extended between $Q$ and $\bar{Q}$ as obtained from the general solution \eqref{sol1} of the EOM \eqref{eomF1}
(see Figure \ref{fig:stringprobe} for an illustration of this configuration). Given a value of the charge parameter $\kappa$, the asymptotically measured temperature $T$ and the distance $L$ between $Q$ and $\bar{Q}$ one can examine whether \eqref{sol1} gives rise to a solution. As we now describe, there can be either zero, one or two solutions available given values of $\kappa$ and $LT$. For a particular solution, one has a particular value of $\sigma_0$ which measures the closest proximity of the F-string probe to the black hole horizon in the background \eqref{adsbhmet}, namely at the point $(z,x)=(\sigma_0,L/2)$. Fixing the value of $\kappa$, a convenient way to examine the available solutions is to turn things around and ask for the available solutions given instead $T$ and $\sigma_0$, or, more precisely, the dimensionless quantity $\hat{\sigma}_0 = \pi T \sigma_0$, which measures the location of the closest proximity of the F-string probe in a rescaled coordinate system where the F-string touching the event horizon would correspond to $\hat{\sigma}_0=1$.  Given $\hat{\sigma}_0$, $LT$ is found from \eqref{LTeq}. One can now plot $LT$ as a function of $\hat{\sigma}_0$ for various values of the charge parameter $\kappa$. This is done numerically in Figure \ref{fig:LTsigmaplots}.

From Figure \ref{fig:LTsigmaplots} we see that for a given $\kappa$ there exists solutions in the range from $\hat{\sigma}_0 =0$ to $\hat{\sigma}_0 = \hat{\sigma}_c ( \kappa)$. Here $\hat{\sigma}_c(\kappa)$ is given by \eqref{hatsigmac} and is the critical distance from the event horizon for which the F-string probe reaches the maximal temperature (note that for small $\kappa$ the probe approximation breaks down before reaching $\hat{\sigma}_c ( \kappa)$). For a given $\kappa$ we write $(LT)_c$ as the value of $LT$ for the solution with $\hat{\sigma}_0 = \hat{\sigma}_c ( \kappa)$. We notice that the quantity $LT$ always has a maximum value $(LT)_{\rm max} $ for a given value of $\kappa$. We denote the
value of $\hat \sigma_0$ for which this maximum is attained by $\hat \sigma_{0,{\rm max}}$.  

If we consider what happens for given values of $\kappa$ and $LT$, we see that for  $LT < (L T)_c$ we find one solution. Instead in the range $(L T)_c \leq LT < (L T)_{\rm max} $ we have two branches of solutions. 
Then for $LT = (LT)_{\rm max} $ we reach the maximal possible value of $LT$ for a given value of $\kappa$. Thus, for $LT > (LT)_{\rm max} $ we do not have any available solutions corresponding to the F-string probe stretching between $Q$ and $\bar{Q}$. 

For comparison we have also plotted in  Figure \ref{fig:LTsigmaplots} the result 
for $LT$  obtained using  the extremal F-string probe. This corresponds  to $\kappa=0$, as can be
seen from \eqref{coshalphaexp} and \eqref{xprimeexp}, and integrating \eqref{xprimeexp} in that case
one finds 
\begin{equation}\label{LTsmallkap}
	LT \vert_{ \kappa =0} =\frac{2 \sqrt{2\pi} }{  \Gamma
   \left(\frac{1}{4}\right)^2}   \hat{\sigma}_0 \sqrt{  1 - \hat{\sigma}_0^4  }  \, \, 
   _2F_1\left(\frac{1}{2},\frac{3}{4};\frac{5}{4};\hat{\sigma}_0^4\right)
   \end{equation}
This matches with the result found in Refs.~\cite{Brandhuber:1998bs} 
using the extremal F-string probe in the  $\ads$ Poincar\'e black hole background.  
 $LT$ also exhibits a maximum in this case: Its value is 
$L_{\rm max} T  \simeq 0.277$ and it is reached for $\hat{\sigma}_0 \simeq 0.85$.  

The thermal F-string probe induces both qualitative and quantitative differences compared to the extremal
F-string probe. The qualitative difference consists in that, while for the extremal probe one has two solutions
available for given $LT$, for the thermal F-string probe there is only one solution when $LT$ is sufficiently small.
This is clear from Figure \ref{fig:LTsigmaplots}. In detail, we see that there is only one solution available for
$LT < (LT)_c$. Taking into account the validity of the probe approximation the values of $LT$ for which there
are two solutions are even more restricted. We also notice a clear quantitative difference
in the dependence of $LT$ on $\hat{\sigma}_0$ as well as the value of $(LT)_{\rm max}$. 
In particular, notice from \eqref{xprimeexp} that the value of $(LT)_{\rm max}$ receives a ${\cal{O}}(\sqrt{\kappa})$ correction for small  $\kappa$, which is thus a $1/N$ effect that is missed by the less accurate extremal F-string probe.  

\begin{figure}[h!]
\centerline{\includegraphics[scale=1.2]{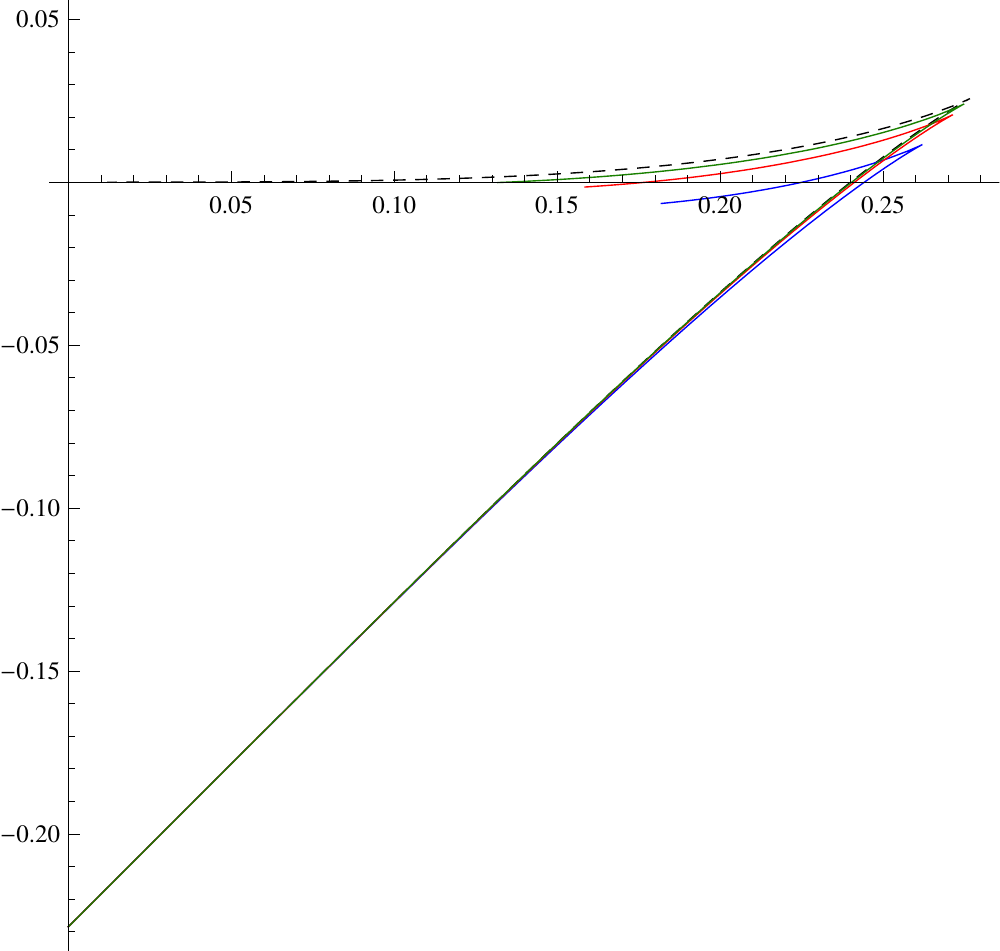}}
\begin{picture}(0,0)(0,0)
\put(410,278){\Large $LT$}
\put(20,335){\Large $ \frac{ L \Delta \CF }{k \sqrt{\lambda}} $}
\end{picture}		
	\caption{The quantity $L \Delta \CF/( k \sqrt{\lambda})$, with $\Delta \CF $ the free energy difference, 
	as a function of  $LT$ for
	various values of $\kappa$: $\kappa=0.01$ (blue line), $\kappa=0.001$ (red line), $\kappa=0.0001$ (green line).
	For comparison we also plot the 
result obtained using the extremal probe (dashed line) \cite{Brandhuber:1998bs} (see Eq.~\eqref{DFsmallkap}). }
	\label{fig:FreeEnQ2comp}
\end{figure}

\begin{figure}[h!]
\centerline{\includegraphics[scale=1.2]{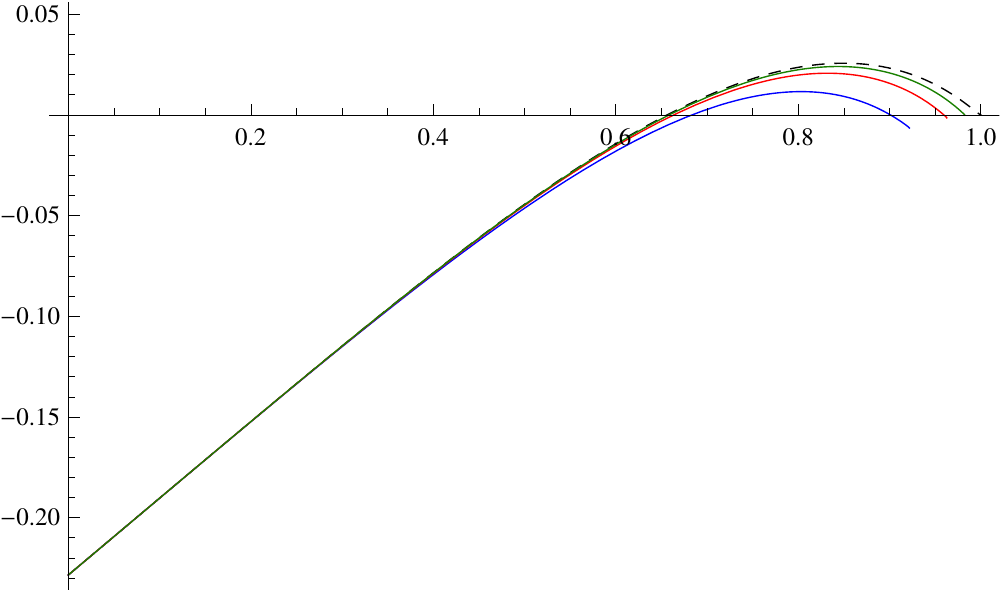}}
\begin{picture}(0,0)(0,0)
\put(410,178){\Large $\hat{\sigma_0}$}
\put(20,212){\Large $ \frac{ L \Delta \CF }{k \sqrt{\lambda}} $}
\end{picture}		
	\caption{The quantity $L \Delta \CF/( k \sqrt{\lambda})$, with $\Delta \CF $ the free energy difference, 
	as a function of $\hat{\sigma}_0$ for
	various values of $\kappa$: $\kappa=0.01$ (blue line), $\kappa=0.001$ (red line), $\kappa=0.0001$ (green line).	For comparison we also plot the 
result obtained using the extremal probe (dashed line) \cite{Brandhuber:1998bs} (see Eq.~\eqref{DFsmallkap}). }
	\label{fig:FreeEnQ1comp}
\end{figure}

We now turn to the free energy of the Wilson loop and its comparison to that of 
two Polyakov loops. As described above, the latter configuration corresponds to two straight strings
stretching towards the horizon from the charge $Q$ and $\bar{Q}$ respectively, and exists for all values of $LT$.
We therefore want to determine which of the phases is preferred by computing the difference in their free energies $\Delta {\cal{F}}$. As we found above, $\Delta \CF = \CF_{\rm loop} - 2\CF_{\rm charge}$ with $\CF_{\rm loop}$ given by \eqref{Floop} and $\CF_{\rm charge}$ by \eqref{Fcharge2}.
In case $\Delta \CF$ is less than zero the Wilson loop is thermodynamically preferred in the canonical
ensemble.  Using the exact solution \eqref{sol1} as well as \eqref{kappa} we have plotted  in Figures \ref{fig:FreeEnQ2comp} and \ref{fig:FreeEnQ1comp}
 for various values of $\kappa$ the quantity $L\Delta {\cal{F}}/(k \sqrt{\lambda})$ as a function of  $LT$
 and $\hat\sigma_0$  respectively.


As in Fig.\ref{fig:LTsigmaplots}, we have also plotted the corresponding results obtained using 
the extremal F-string probe. This can be found again by  substituting the $\kappa =0$ limit of 
\eqref{coshalphaexp}  into the expressions \eqref{Floop}, \eqref{Fcharge2}, yielding for the
free energy difference \eqref{DeltaF} 
\begin{align}\label{DFsmallkap}
	\Delta {\cal{F}}  \vert_{\kappa=0} &= \sqrt{\lambda} kT \left[   -\frac{1}{\hat{\sigma}_0} + 1 + 
	\int_0^{\hat{\sigma}_0} d\hat{\sigma}\frac{1}{\hat{\sigma}}
	\left(\hat{\sigma}_0^2\sqrt{\frac{\hat{\sigma}^4-1}{\hat{\sigma}^4- \hat{\sigma}_0^4}}-1\right)  \right]  \nonumber \\
	 &= \sqrt{\lambda} kT \left[ 1 - 
	\frac{\sqrt{2} \pi^{3/2}  \left(1 - \hat{\sigma }_0^4\right)}{\hat{\sigma }_0 \Gamma \left(\frac{1}{4}\right)^2}\,
	 _2F_1\left(\frac{1}{2},\frac{3}{4};\frac{1}{4};\hat{\sigma
   }_0^4\right) \right]  \,,
\end{align}
This agrees with expression  of the energy of the solution found in Refs.~\cite{Brandhuber:1998bs}.

Focussing first on Figure \ref{fig:FreeEnQ2comp}, we see that the intersection of the curves with the horizontal axis  determines the onset, denoted by  $(LT)\vert_{\Delta \CF =0}$,  where the thermodynamically
preferred configuration is that of two straight strings as illustrated in Figure \ref{fig:stringprobe} corresponding to two Polyakov loops in the gauge theory.  The corresponding phase transition,
also observed for the extremal probe, can be interpreted as Debye screening of the $Q$-$\bar{Q}$
pair. One finds that, as one moves to (small) non-zero values of $\kappa$, the onset moves to higher values of $LT$. We thus see that using the thermal F-string probe we find that the pair is less easily screened compared to what one can see with the less accurate extremal F-string probe, and that this effect moreover becomes stronger as  $\kappa$ increases.  
Note also that one never reaches $(LT)_{\rm max} $ as the quarks are screened before reaching this distance. 
Finally, we note that the intersection of the curves with the vertical
axis corresponds to the leading Coulomb potential term given in \eqref{FloopLT}.

Another aspect that one can see from Figure  \ref{fig:FreeEnQ2comp} is
the free energy comparison for the two phases of Wilson loops  that are present for certain values of $LT$. 
The cusp of the swallow tail is where one reaches  $(LT)_{\rm max} $,  with the lower part of the swallow tail 
corresponding to $0< \hat \sigma_0  < \hat \sigma_{0,{\rm max}}$, and
the upper part corresponding to  $\hat \sigma_{0,{\rm max}} \leq \hat \sigma_0 \leq \hat \sigma_c$. 

The features mentioned in regard to Figure \ref{fig:FreeEnQ2comp} can alternatively be seen from Figure \ref{fig:FreeEnQ1comp}, now as function of $\hat{\sigma}_0$. Denoting $\hat \sigma_0 \vert_{\Delta \CF =0}$
as the onset for quark screening, one may check in this connection 
that  $\hat \sigma_0 \vert_{\Delta \CF =0} < \hat \sigma_{0,{\rm max}} < \hat \sigma_c $
is valid for any value of $\kappa$.

We thus conclude that, with respect to the free energy and in particular quark screening, 
 the thermal F-string probe sees quantitative effects that the less accurate extremal F-string probe misses. 
In particular, for the onset of quark screening we find for small $\kappa$ 
\begin{equation}
(LT)\vert_{\Delta \CF =0} \simeq 
0.240038 + 0.0379706 \sqrt{\kappa}  
\end{equation}
so a $\sqrt{\kappa}$ correction term to the  result obtained previously using the extremal F-string probe. 

Thus for a finite value of $LT$, the quantitative difference mainly consists in having $\sqrt{\kappa}$ corrections to the extremal probe results. From the point of view of the gauge theory this corresponds to a $\sqrt{k} \lambda^{1/4} /N$ correction to the previously obtained results for the potential between the charges, as well as to the critical value
 $(L T)\vert_{\Delta \CF =0}  $ of $LT$ where the charges becomes screened. 

Finally, we note that the thermal effects observed in the system studied in this paper, bear some resemblance
to those found for the thermal BIon studied in Refs.~\cite{Grignani:2010xm,Grignani:2011mr}. In that case, the blackfold approach was used to study a thermal D3-F1 brane probe in  hot  flat space.  This provides a 
description of the finite temperature generalization of the BIon configuration consisting of a D-brane and a parallel anti-D-brane connected by a wormhole with F-string charge. Also in that case one finds that the finite temperature system behaves qualitatively different compared to its zero-temperature counterpart. In particular, for a given separation between the D-brane and anti-D-brane
there are either one or three phases available, while at zero temperature there are two phases.
Furthermore, analysis of the free energy of the finite temperature generalization of the BIon shows a similar swallow tail structure as found above.

\section{Discussion}
\label{sec:concl}

We conclude with a brief discussion of our results and open problems. Our analysis has been the first concrete application
of the blackfold method
\cite{Emparan:2009cs,Emparan:2009at,Emparan:2011hg,Caldarelli:2010xz,Grignani:2010xm}
 in the context of the AdS/CFT correspondence. As explained in \cite{Grignani:2010xm} 
(see also \cite{Emparan:2011hg})  the method provides a new
tool for thermal probes in finite temperature backgrounds, taking into account the fact that for extended probes the internal
degrees of freedom should be in thermal equilibrium with the background. The application of this method in this paper
for the study of Wilson loops in finite temperature $\CN=4$ SYM using thermal F-string probes in the $\ads$
black hole background, confirms that there are both qualitative and quantitative effects that the less accurate extremal F-string probe misses.  Consequently, it seems
relevant to apply this new computational tool to other settings and setups, notably in the AdS/CFT correspondence. 
As one direction, it would be very interesting to generalize our results to heavy quarks by introducing appropriate boundary conditions for the string close to the boundary of $\ads$. One could then also examine, using this new perspective, other holographic aspects of quark-gluon plasma physics, such as the energy loss of a heavy quark moving through the plasma (see e.g. the review 
\cite{Chernicoff:2011xv} and references therein). 

Furthermore, one could use the method to revisit the thermal generalization of  the Wilson loop in higher representations,
in the regime where it involves a ``blown-up'' version using a D3-brane  \cite{Drukker:2005kx} (symmetric representation) or a D5-brane \cite{Yamaguchi:2006tq} (antisymmetric representation). In particular, this may be interesting in view of
the discrepancies between gauge theory and gravity results found in~\cite{Hartnoll:2006hr}
for the symmetric representation. This would involve considering a SUGRA  black D3-brane probe in the $\ads$ black hole background. 

For the case studied in this paper, one of the novel effects is the new term in the free energy for 
the rectangular Wilson loop, in the small $LT$ regime.  Relative to the leading Coulomb potential, this term is proportional to $ \sqrt{k} \lambda^{1/4} (LT)^3/N$.  It would be very interesting to examine whether 
such a correction also appears at weak 't Hooft coupling in ${\cal{N}}=4$ SYM at finite temperature. 

It is important to mention that just as we in this paper have considered the thermal backreaction effects in 
the worldvolume of the string, one can also consider the  backreaction on the background \eqref{adsbhmet}
due to the thermal radiation in the bulk. Unlike the effects found in this paper, which are of order $1/N$, the
bulk backreaction is of order $1/N^2$,  since it is governed by the gravitational coupling, and is therefore subleading.

Finally, we note that in the construction of this paper (as well the previous works \cite{Brandhuber:1998bs}
that use the extremal F-string probe) the F-string is localized on the five-sphere. One may wonder if there is a competing configuration
in which the probe is smeared over the five-sphere, but such configuration  would suffer from a Gregory-Laflamme type instability.
We also note that the blackfold approach provides a powerful tool to study time evolution and stability
 \cite{Emparan:2009at,Camps:2010br} and that corrections that go beyond the thin brane limit have been addressed recently
 in \cite{Armas:2011uf}.  It would be interesting to use these insights to further study the case considered in this paper.

\section*{Acknowledgments}

We thank Jay Armas, Francesco Bigazzi, Jan de Boer, Aldo Cotrone, Roberto Emparan, Carlos Hoyos and Gordon Semenoff 
for useful discussions. TH thanks NBI and NO thanks Nordita for hospitality. 
GG, TH,  NO and MO are grateful to the GGI in Firenze for hospitality
during the workshop on ``Large $N$ Gauge Theories''.
This work was supported in part by the MIUR-PRIN contract 2009-KHZKRX.
The work of NO is supported in part by the Danish National Research Foundation project ``Black holes and their role in quantum gravity''.


\addcontentsline{toc}{section}{References}

\small



\providecommand{\href}[2]{#2}\begingroup\raggedright\endgroup

\end{document}